\documentclass[%
 reprint,
 amsmath,amssymb,
 aps,prb
]{revtex4-1}

\usepackage{graphicx}
\usepackage{dcolumn}
\usepackage{bm}

\usepackage{xcolor}
\begin{document}
\title{Josephson effect in type-I Weyl Semimetals}
\author{Debabrata Sinha}
\affiliation{Center for Theoretical Studies, Indian Institute of Technology,
Kharagpur-721302, India}
\date{\today}
\begin{abstract}
The emergent Weyl fermions in condensed matter generally break the Lorentz invariance resulting in a tilted (type-I) or over-tilted (type-II) energy dispersion. The tilting energy spectrums can lead to exotic quantum interference effects in a junction set up. Here, we theoretically investigate the Josephson current in a Weyl superconductor-Weyl (semi)metal-Weyl superconductor junction of a time-reversal (TR) broken type-I Weyl semimetal. We demonstrate that the Cooper pairs of BCS-like pairing acquire a finite momentum in case of inversion symmetric tilt. Consequently, the system exhibits tilt induced anomalous current phase relations which are manifested by supercurrent $0$-$\pi$ transition and Josephson $\phi$ junction. On the contrary, these effects remain absent in case of inversion breaking tilt and for FFLO-like pairing in the Weyl superconductor. We further chart out qualitative differences between the two distinct types of pairings by studying the critical current dependency on junction length. Our study opens a new avenue to probe the unconventional superconducting pairings in TR-broken Weyl semimetals. It is also quite interesting that the tilting in Weyl nodes naturally leads to anomalous current phase relations in this model without any magnetic manipulation!
\end{abstract}
\maketitle
\section{Introduction}
A Weyl semimetal (WSM) hosts three-dimensional gapless topological states emanated from $k$-space singularities. The k-space singularities persist due to the merging of valance and conduction band at some specific k-points in the Brillouin Zone, known as Weyl nodes. The nodes always appear in pairs, carry opposite topological charges, and are protected due to either time-reversal (TR) or inversion (IR) symmetry. The energy-momentum dispersion is linear and therefore, Weyl cones are formed around the nodes. 

The realization of these gapless topologically nontrivial states have drawn much attentions\cite{Armitage-RMP18,Burkov-Ann18}. WSMs were predicted theoretically\cite{Burkov-PRL11,Wan-PRB11,Xu-PRL11} and observed experimentally in a wide range of materials \cite{Xu-Science15,Xu-Sci. Adv15, Hasan-Science15,Yang-Nat15}. Most of the experimental findings Weyl fermions have anisotropic and tilted (type-I) or over tilted (type-II) energy dispersion\cite{Bernevig-Nat15,Zhang-Nat-Com17}. In type-I, the Weyl cone is weakly tilted and the Fermi surface is a point like at the nodes. In type-II, the tilt exceeds the Fermi velocity of an electron/hole and generates electron-hole pockets near the Weyl nodes. The tilt violates the Lorentz invariance which is a fundamental symmetry of Weyl fermions in high energy physics. The violation of Lorentz symmetry is quite natural in the condensed matter since the velocity of quasiparticles always less than the velocity of light. However, the tilt in the Weyl cones does not alter the topology of energy bands. Rather, it largely affects the quantum transport \cite{Max-PRB15} including Klein tunneling \cite{Nguyen-PRB18,Beenakker-PRL16}, spin transport \cite{Sinha-EPJB19}, Andreev reflection \cite{Hou-PRB17,Jafari-PRB19}, magnetotransport\cite{Sinha-arXiv}. 

The valance and conduction bands remain filled and empty respectively in a semimetal. The Fermi level remains situated at their touching point i.e., at the Weyl nodes. Impurities cause the Fermi level to enter into the conduction or valance band resulting in a finite density of states near Weyl nodes. The tilting can cause non zero density of states at the nodes even in the absence of impurities. These naturally motivate to query about the superconducting states in Weyl metals. Moreover, the non-trivial topology and nondegenerate valance and conduction bands may trigger unconventional superconducting states in Weyl metals.

Many works have been devoted to understanding the superconducting pairing mechanisms of Weyl metals\cite{Wang-ScB,Kim-PRB16,Moore-PRB12,Wang-PRB16,Aji-PRB14,Bednik-PRB15}. Two distinct types of cooper pairings were predicted\cite{Moore-PRB12,Aji-PRB14,Bednik-PRB15}: an even/odd parity BCS(Bardeen-Cooper-Schrieffer)-like pairing with zero cooper pair momentum and a FFLO-like pairing with finite-momentum pairs. In BCS state, the electron at momentum $\mathbf{k}$ near one Weyl node pairs with opposite chiral Weyl node electron at momenta $-\mathbf{k}$ (internode) with same energy (let assume TR-symmetry is broken but IR-symmetry is preserved) whereas, in FFLO state, pairs are formed from the same chiral Weyl node (intranode).  The FFLO state is only pairing term when both IR and TR symmetry is broken. In this situation, two opposite chiral Weyl nodes are shifted to different energy values. The mean-field calculation predicts that local phonon-mediated attractive interaction favors finite momentum FFLO-like pairing over the even-parity BCS state\cite{Moore-PRB12}. In contrast, Ref.\cite{Aji-PRB14} predicts that non-local interaction endorses odd parity BCS state over FFLO state and the BCS state vanishes identically for local interactions. The odd parity BCS ground state is also predicted in inversion symmetric WSMs\cite{Bednik-PRB15}.  However, further experiments are required to understand the proper superconducting mechanisms in Weyl metals.

The Andreev reflection and Josephson effect are basic tools to investigate the unconventional superconducting pairings. Recently, several works have been reported in order to understand the proper superconducting mechanisms in a WSM \cite{Xing-EPL13,Madsen-PRB17,Ma-PRB18}. Ref.\cite{Madsen-PRB17} shows that the Josephson effect for FFLO-like pairing of Weyl SNS junction has closely resembled the theory of Josephson effect of graphene or topological insulators. On the other hand, the critical current is independent of chemical potential for BCS-like pairing and thus the effect is different from the FFLO state or other two dimensional topological materials. However, the tilting in Weyl nodes is ubiquitous and a tilting spectrum always favors the onset of superconductivity. This prompts us to study the tilting effect in the Josephson current. Most importantly, we explore whether the tilt induced Josephson effect can be used as a tool to distinguish the distinct types of pairing mechanisms in these scenarios.

The ground state of a Josephson junction has a sinusoidal variation with the superconducting phase difference $\phi$, noticed in most of the experimental junctions for a long time been\cite{Josephson}. However, the development of the fabrication technique enables us to detect a large variety of current phase relations (CPRs). For example, Josephson $\pi$ junction with free energy ground state at $\phi=\pi$ occurs in diverse physical systems. These include superconductor -ferromagnet-superconductor Josephson junction\cite{Ryazanov-PRL01,Buzdin-RMP05}, Josephson junction with unconventional superconducting order parameter \cite{Tsuei-RMP,Lombardi-PRL02}, Josephson junctions based on topological insulators or nanowire in presence of Zeeman field\cite{Hart-Nat17, Murani-Nat17,Yokoyama-PRB14}, Josephson junction of an irradiated Weyl semimetal \cite{Khanna-PRB17} and strong spin-orbit coupled two dimensional materials\cite{Zhou-PRB16}. A chirality imbalanced potential also leads $0$-$\pi$ transition for BCS-like pairing of a TR broken WSM\cite{Uddin-PRB19}. The anomalous supercurrent can flow in $\phi$ state with a free energy ground state other than $0$ or $\pi$\cite{Sickinger-PRL12,Szombati-Nature}. The corresponding CPR in this situation is given: $J=J_c\sin(\phi-\phi_0)$, known as Josephson $\phi$ junction which lacks the phase inversion symmetry. Nonsinusoidal CPR with higher-order harmonic terms has been reported in a variety of topological materials\cite{English-PRB16,Sochnikov-PRL15,Spanton-Nat17}. In Refs.\cite{Alidoust-PRB18,Alidoust-PRB20}, the supercurrent reversal and Josephson $\phi$ junction were reported by tuning the magnetic field or other relevant parameters in a tilted Weyl Hamiltonian. 

In the present work, we study the Josephson effect in a TR-broken type-I WSM. We consider both FFLO and BCS-like pairings in the Weyl superconductor. We demonstrate that for IR symmetric tilt (i.e., the opposite chiral cones are tilted in the opposite direction), the phase relation of Andreev bound state (ABS) with $\phi$ provokes entirely different signatures for the two pairing mechanisms. ABS spectrums from two chiral Weyl nodes are always degenerate for FFLO-like pairing. The spectrums are degenerate for BCS-like pairing at specific values of tilt induced phase $\phi_t$ with the different realization of the ground state. Our study reveals that the phase $\phi_t$ has an opposite sign at opposite chiral Weyl nodes for BCS-like pairing. This leads to several peculiar phenomena in the CPRs including the supercurrent reversal and Josephson $\phi$ junction. At supercurrent, $0$ to $\pi$ transition the current phase relation is dominated by second harmonic. These anomalous CPRs are absent for the FFLO pairing and for IR symmetry breaking tilt. In these states, $\phi_t$ has the same sign at opposite chiral nodes which has a trivial effect in CPRs. The phase $\phi_t$ is tunable by the junction length and doping. We discuss the critical current dependencies on the length of the normal Weyl metal region and anticipate the qualitative differences between the two pairings. These provide a route to distinguish between the distinct types of pairing states in TR-broken tilted Weyl semimetals.

This paper is organized as follows. In Sec-II we analyze the theory of FFLO and BCS pairing and discuss our model. In Sec-III, the basic formulas for Andreev bound states and Josephson current are constructed. In Sec-IV, we discuss the results and finally in Sec-V the conclusion of this work is given.

\section{Theory}
We consider Josephson junction made of type-I WSM with a slab of normal type-I Weyl (semi)metal for $0<z<L$ is sandwiched between two type-I heavily doped Weyl superconducting regions. The left and right superconducting regions extend semi-infinetly along $\hat{z}$-direction. We consider TR-broken WSMs with minimum two opposite chiral Weyl nodes, which are situated at $\pm \mathbf{K}_0$ on $q_x-q_z$ plane with $\mathbf{K}_0=K_0(\mathbf{e_z}\cos\alpha+\mathbf{e_x}\sin\alpha)$. Here, $\alpha$ is the angle between crystal coordinates and junction coordinates. The normal-state two band Hamiltonian with momenta $\mathbf{k}=\pm \mathbf{K}_0+\mathbf{q}$ around the Weyl nodes at $\pm \mathbf{K}_0$ reads \cite{Xing-EPL13,Madsen-PRB17,Breunig-PRB19},
\begin{eqnarray}
H_0=\sum_{\chi}\sum_{\mathbf{q}}\Psi^\dagger_{\chi}(\mathbf{q})h^W_\chi\Psi_{\chi}(\mathbf{q})
\end{eqnarray}
with
\begin{eqnarray}
h^W_{\chi}&=& \hbar(a_1 q_1+a_3 q_3)\sigma_0+ \hbar v(q_1\sigma_1+q_2\sigma_2+\chi q_3\sigma_3)-\mu\nonumber\\
&=&h_t +h_{\chi}
\label{tilt-Hamil}
\end{eqnarray}
Here, $\chi=\pm$ defines the chirality of the Weyl nodes. The first term in Eq.(\ref{tilt-Hamil}), $h_{t}= \hbar(a_1 q_1+a_3 q_3)\sigma_0$ is responsible for tilting in the Weyl nodes. For simplicity, we consider tilting is along $q_1$ and $q_3$ direction with strength $a_1$ and $a_3$ respectively. $\sigma_0$ and $\sigma_i$'s are unit and Pauli matrices acting on the spin space, respectively. $\Psi^\dagger_{\chi}(\mathbf{q})=(c^\dagger_{\uparrow,\chi}(\mathbf{q}),c^\dagger_{\downarrow,\chi}(\mathbf{q}))$ is the spinor basis with $c^\dagger_{\sigma,\chi}(\mathbf{q})$ the creation operator for an electron. The two different coordinates systems are related as follows, 
\begin{eqnarray}
q_1&=&q_x\cos\alpha-q_z\sin\alpha\nonumber\\
q_2&=&q_y\nonumber\\
q_3&=&q_z\cos\alpha+q_x\sin\alpha
\end{eqnarray}
and, similarly, $\sigma_1=\sigma_x\cos\alpha-\sigma_z\sin\alpha$, $\sigma_2=\sigma_y$, $\sigma_3=\sigma_z\cos\alpha+\sigma_x\sin\alpha$. The pairing term for BCS and FFLO pairings are given\cite{Xing-EPL13},
\begin{eqnarray}
\mathcal{H}^{B}_{pair}&=&\sum_{\chi,\mathbf{q}}\Delta(z)c^\dagger_{\uparrow,\chi}(\mathbf{q})c^\dagger_{\downarrow,-\chi}(-\mathbf{q})+h.c.\nonumber\\
\mathcal{H}^{F}_{pair}&=&\sum_{\chi,\mathbf{q}}\Delta(z)c^\dagger_{\uparrow,\chi}(\mathbf{q})c^\dagger_{\downarrow,-\chi}(\mathbf{q})+h.c.
\end{eqnarray}
where the subscript $B$ and $F$ correspond to BCS anfd FFLO-like pairing, respectively. $\Delta(z)$ is the pairing potential. The BdG Hamiltonian in the basis of $(c^\dagger_{\uparrow,+}(\mathbf{q}),c^\dagger_{\downarrow,+}(\mathbf{q}),c_{\downarrow,-}(-\mathbf{q}),-c_{\uparrow,-}(-\mathbf{q}))$ and $(c^\dagger_{\uparrow,-}(\mathbf{q}),c^\dagger_{\downarrow,-}(\mathbf{q}),c_{\downarrow,+}(-\mathbf{q}),-c_{\uparrow,+}(-\mathbf{q}))$ are given, both BCS and FFLO-like pairings,
\begin{eqnarray}
H^{\pm}_{B}=
\begin{pmatrix}
h^W_{\pm}(-i\mathbf{\nabla}\mp\mathbf{K}_0) & \Delta(z)\\
\Delta(z)^{*} & -h^W_{\mp}(-i\mathbf{\nabla}\mp\mathbf{K}_0)
\label{BCS}
\end{pmatrix}\\
H^{\pm}_{F}=
\begin{pmatrix}
h^W_{\pm}(-i\mathbf{\nabla}\mp\mathbf{K}_0) & \Delta(z)e^{\pm2i\mathbf{K}_0\cdot\mathbf{r}}\\
\Delta(z)^{*}e^{\mp2i\mathbf{K}_0\cdot\mathbf{r}} & -h^W_{\pm}(-i\mathbf{\nabla}\pm\mathbf{K}_0)
\label{FFLO}
\end{pmatrix}
\end{eqnarray}
For BCS-like pairing in Eq.(\ref{BCS}), the pairing potential couples the electrons from two opposite chiral nodes whereas for FFLO-like pairing in Eq.(\ref{FFLO}) it couples from the same node. A gauge transformation removes the large momentum $\mathbf{K}_0$ from the BdG Hamiltonian in Eqs.(\ref{BCS},\ref{FFLO}). The transformation for BCS and FFLO pairings are\cite{Madsen-PRB17},
\begin{eqnarray}
H^{\pm}_{B}\rightarrow \tilde{H}^{\pm}_{B}&=&e^{\pm i\mathbf{K}_0\cdot\mathbf{r}}H^{\pm}_{B}e^{\mp i\mathbf{K}_0\cdot\mathbf{r}}\\
H^{\pm}_{F}\rightarrow \tilde{H}^{\pm}_{F}&=&e^{\pm i\sigma_z\mathbf{K}_0\cdot\mathbf{r}}H^{\pm}_{F}e^{\mp i\sigma_z\mathbf{K}_0\cdot\mathbf{r}}
\end{eqnarray}
respctively, which gives the transformed Hamiltonian,
\begin{eqnarray}
\tilde{H}^{\pm}_{B}=\begin{pmatrix}
h^W_{\pm}(-i\nabla) & \Delta(z)\\
\Delta(z)^{*}&-h^W_{\mp}(-i\nabla)
\label{bcs-hamil-gauge}
\end{pmatrix}\\
\tilde{H}^{\pm}_{F}=\begin{pmatrix}
h^W_{\pm}(-i\nabla) & \Delta(z)\\
\Delta(z)^{*}&-h^W_{\pm}(-i\nabla)
\end{pmatrix}
\label{fflo-hamil-gauge}
\end{eqnarray}
The Hamiltonian $h_{+}$ in Eq.(\ref{tilt-Hamil}) is independent of $\alpha$. We perform an extra unitary transformation \cite{Madsen-PRB17,Breunig-PRB19} to remove the angle $\alpha$ from the hole part of the BdG Hamiltonian in Eq.(\ref{bcs-hamil-gauge}), 
\begin{eqnarray}
\tilde{H}^{\pm}_{B}\rightarrow \hat{U}^{\pm}_{\alpha}\tilde{H}^{\pm}_{B}(\hat{U}^{\pm}_{\alpha})^{-1}
\end{eqnarray}
with 
\begin{eqnarray}
\hat{U}^{\pm}_{\alpha}=\frac{1}{2}[(\tau_0\pm\tau_z)\sigma_x e^{i\alpha\sigma_y}+(\tau_0\mp\tau_z)]
\end{eqnarray}
The unit matrix $\tau_0$ and Pauli matrix $\tau_i$ are acting on particle-hole space.
The resulting BdG Hamiltonian become,
\begin{eqnarray}
\tilde{H}^{+}_{B}=\begin{pmatrix}
H^e_{B}&\tilde{\Delta}(z)\\
\tilde{\Delta}(z)^{*}& H^h_{B}
\end{pmatrix}
\label{bcs-final-hamil}
\end{eqnarray}
with $H^e_{B}=h_t+h_{+}(-i\nabla)$ and $H^h_{B}=-h_t-\bar{h}_{+}(-i\nabla)$. Similarly, we can write down the Hamiltonian for $\tilde{H}^{-}_{B}$. The $\alpha$ dependency shifted in the pair potential and modified form is: $\tilde{\Delta}(z)=\Delta(z)\sigma_x\cos\alpha-\Delta(z)\sigma_z\sin\alpha \approx -\Delta(z)\sigma_z\sin\alpha$\cite{Madsen-PRB17}. Here, $\bar{h}_{+}=\hbar v(q_x\sigma_x-q_y\sigma_y+q_z\sigma_z)$. The tilting part of the Hamiltonian is: $h_t=\hbar (a_1\cos\alpha+a_3\sin\alpha)q_x+(-a_1\sin\alpha+a_3\cos\alpha)q_z$. In the rest of the paper we take $\alpha=\pi/2$ and consider tilting is only along the transport direction ($z$-axis) i.e., $h_t=\hbar C_{\chi}q_z$.

\begin{figure}
\includegraphics[width=1.67in]{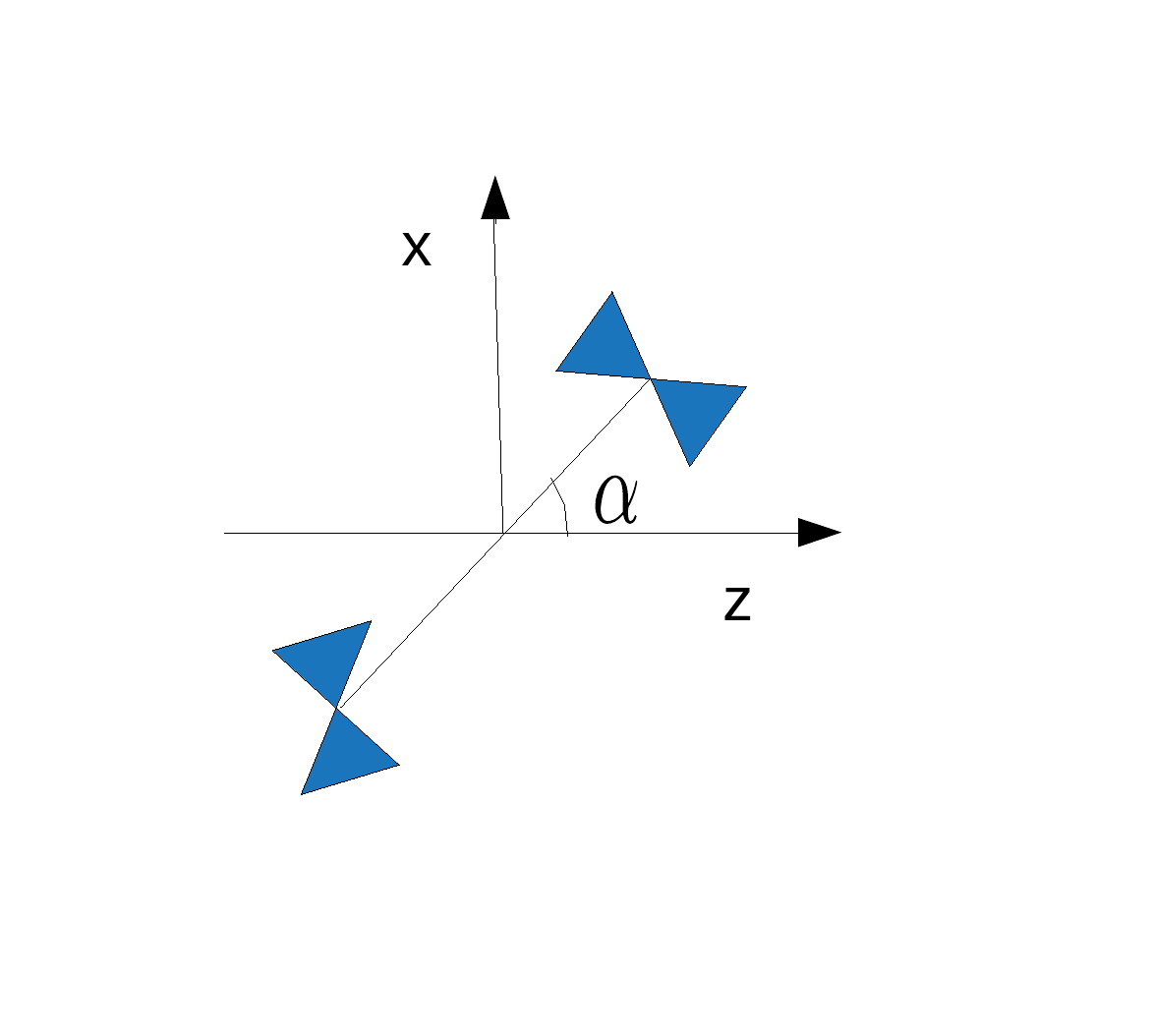}
\includegraphics[width=1.67in]{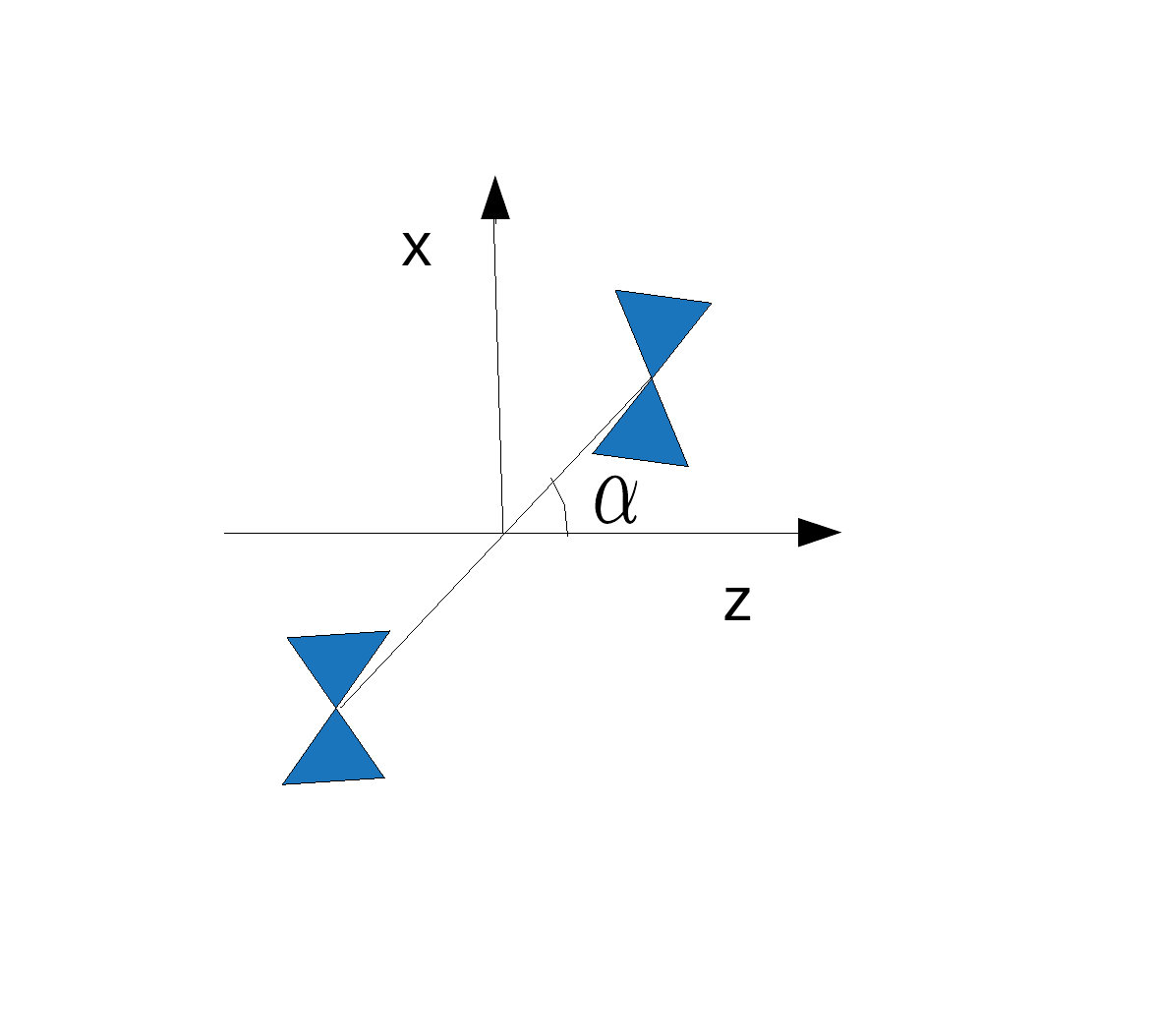}
\caption{Schematic diagram: The left panel show the inversion symmetric tilt of the Weyl nodes (Case-I in the text). In this case, the opposite chiral Weyl nodes are tilted in opposite direction. The right panel show the inversion breaking tilt of the Weyl nodes (Case-II in the text). In this case, the opposite chiral Weyl nodes are tilted in the same direction.}
\hspace{3pt}
\label{sch-fig}
\end{figure}

A WSM is in type-I phase if $C_\chi <v$ and in type-II phase if $C_\chi > v$. The model Hamiltonian in Eq.(\ref{tilt-Hamil}) is inversion symmetric (i.e., $\sigma_3 h^W_+(\mathbf{q})\sigma_3=h^W_{-}(-\mathbf{q})$) if $C_{+}=-C_{-}$ (Case-I). In this case, the opposite chiral Weyl nodes have tilts in opposite direction (See Fig.(\ref{sch-fig})). The inversion symmetry is broken if $C_{+}=C_{-}$ (Case-II). In this case, the opposite chiral Weyl nodes tilts in the same direction (See Fig.(\ref{sch-fig})). Following Ref.\cite{Bednik-PRB15}, we emphasize that the BCS pairing is the dominant pairing in Case-I and the FFLO is the dominat pairing in Case-II.  However, we discuss both tilting cases in the Josephson current of a type-I WSM. In the Josephson junction, we assume a steplike model for $\Delta(z)$ and $\mu$. We consider pairing potential: $\Delta(z)=|\Delta|[\Theta(-z)e^{i\phi/2}+\Theta(z-L)e^{-i\phi/2}]$ with $|\Delta|$ is the superconducting gap and $\phi$ is the phase difference of superconducting order parameter. The chemical potential is given: $\mu=\mu_N \Theta(L-|z|)+\mu_S \Theta(|z|-L)$. We take $\hbar=v=1$ and put them back when necessary.

\section{Andreev Bound state and Josephson current}
The Josephson current in the junction is obtained by claculating the Andreev-bound state in the normal region. This is done by matching the wave functions at the interface between three different regions. Explicitly, the wave functions in three different regions are given, 
\begin{eqnarray}
\Psi^{L}_S &=& t_1 \Psi_1 +t_2 \Psi_2\nonumber\\
\Psi_N &=& a_1 \Psi^e_{+}+a_2\Psi^e_{-}+ a_3 \Psi^{h}_{+}+a_4 \Psi^{h}_{-}\nonumber\\
\Psi^{R}_{S} &=& t_3 \Psi_3 +t_4 \Psi_4\nonumber\\
\end{eqnarray}
Here, $\Psi^{L(R)}_S$ is the wave function in the left (right) superconducting region and $\Psi_N$ is the wave function in the normal region. $t_i$'s and $a_i$'s are the scattering coefficients of quasiparticles (electron or hole) in different regions. The subscript $\pm$ on the wave function in the normal region indicates the direction of quasiparticles motion (group velocity). We now look for an energy eigenvalues $\epsilon$ which gives a non zero solution for the boundary conditions: $\Psi^{L}_{S}=\Psi_N$ at $z=0$ and $\Psi_{N}=\Psi^{R}_{S}$ at $z=L$. These boundary conditions leads to $8\times 8$ matrix $\mathcal{M}$ \cite{Linder-PRB09,Sinha-ACP18,Sinha-PRB}:
\begin{eqnarray}
\mathcal{M}=\begin{pmatrix} \mathcal{M}_1 & \mathcal{M}_2\\ \mathcal{M}_3 & \mathcal{M}_4 \end{pmatrix}
\end{eqnarray}
where every elements $\mathcal{M}_i$ are the $4\times 4$ matrix. $Det[\mathcal{M}]_{\epsilon_b}=0$ gives the non-trivial relation between $\epsilon_b$ and superconducting phase difference $\phi$. It is known that, the Josephson current at low temperature ($T \ll \Delta_0/k_b$, with $k_b$ is the Boltzmann constant) is determined solely by the bound states ($\epsilon_b$) and is given by,
\begin{eqnarray}
I(\phi)=-\frac{2e}{\hbar}\sum_{b}{\partial \epsilon_{b}\over \partial \phi} f(\epsilon_b)
\label{Joseph-curr}
\end{eqnarray}
where $f(\epsilon_b)$ is the Fermi-Dirac distribution function. The Josephson current density can obtained,
\begin{eqnarray}
J(\phi)=\frac{W^2}{(2\pi)^2}\int I(\phi) dk_x dk_y
\label{Joseph-dens}
\end{eqnarray}
with $W$ is the dimension in both $x$ and $y$ directions. We define critical supercurrent as $J_c=|max \{J(\phi)\}|$. We take the limit $\mu_N,\mu_s \gg \Delta$. We also consider the short-junction limit i.e., $L\ll \xi=\hbar v/\Delta_0$, which allow us to neglect the Josephson current contributions from the states $\epsilon_b >\Delta_0$. In the following, using this method we calculate Josephson current both in FFLO and BCS-like pairings. Here we focus on zero temperature.


\subsection{FFLO-like Pairing}
We consider the BdG Hamiltonian for FFLO-like pairing given in Eq.(\ref{fflo-hamil-gauge}) and write down the wave functions in three different regions. The electron and hole wave functions in the normal region are follows,
\begin{eqnarray}
\Psi^{e+(-)}_{in(out)}=e^{ik^{+(-)}_{1(2)}z}\begin{pmatrix} 1 & \mathcal{P}^{+(-)}_{1(2)} & 0 & 0 \end{pmatrix}\nonumber\\
\Psi^{h+(-)}_{in(out)}=e^{ik^{+(-)}_{3(4)}z}\begin{pmatrix} 0 & 0 & 1 & \mathcal{P}^{+(-)}_{3(4)} \end{pmatrix}
\end{eqnarray}
Here, $\pm$ sign in uperscript corresponds to the Hamiltonian $H^{\pm}_F$ in Eq.(\ref{fflo-hamil-gauge}) and $in(out)$ denotes the inward and outward particles motion. We skip the wave functions dependence on the transverse coordinates, which is $e^{ik_xx+ik_yy}$, throughout. The elements $\mathcal{P}^{+(-)}_i$ are follows,
\begin{eqnarray}
\mathcal{P}^{+(-)}_1=\frac{k_pe^{i\theta}}{k^{+(-)}_{+}+k^{+(-)}_1};
\mathcal{P}^{+(-)}_2=\frac{k_pe^{i\theta}}{k^{+(-)}_{-}+k^{+(-)}_2}\nonumber\\
\mathcal{P}^{+(-)}_3=\frac{k_pe^{i\theta}}{k'^{+(-)}_{+}+k^{+(-)}_3};
\mathcal{P}^{+(-)}_4=\frac{k_pe^{i\theta}}{k'^{+(-)}_{-}+k^{+(-)}_4}\nonumber\\
\label{we-fflo}
\end{eqnarray}
 where $k_p=\sqrt{k^2_x+k^2_y}$ is the conserved transversed momenta and $\theta=\tan^{-1}(k_y/k_x)$. The wavevectors $k^{+(-)}_i$'s are obtained from the eigenvalues equations of electron and hole Hamiltonian, which are given,
\begin{eqnarray}
k^{+(-)}_{1(2)}&=&\frac{C_{+(-)}(E+\mu_N)\mp \sqrt{(E+\mu_N)^2+(C^2_{+(-)}-1)k^2_p}}{(C^2_{+(-)}-1)}\nonumber\\
k^{+(-)}_{3(4)}&=&\frac{C_{+(-)}(\mu_N -E)\pm \sqrt{(E-\mu_N)^2+(C^2_{+(-)}-1)k^2_p}}{(C^2_{+(-)}-1)}\nonumber\\
\label{wv-fflo}
\end{eqnarray}
with,
\begin{eqnarray*}
k^{+(-)}_{+(-)}&=&\sqrt{{k^{+(-)}}^2_{1(2)}+k^2_p}\nonumber\\
k'^{+(-)}_{+(-)}&=&\sqrt{{k^{+(-)}}^2_{3(4)}+k^2_p}
\end{eqnarray*}
The quasiparticles energy spectrums of BdG hamiltonian are obtained by diagonalizing the Hamiltonian in Eq.(\ref{fflo-hamil-gauge}). The energy eigenvalues are given by,
\begin{eqnarray}
\mathcal{E^{+(-)}}=\pm \sqrt{(\mu_s -C_{+(-)}k_z \pm \mathbf{k})^2+\Delta^2}
\label{Eigen-BdG}
\end{eqnarray}
where $\mathbf{k}=\sqrt{k^2_z+k^2_p}$. We write down the wave functions in the superconducting region with taking consideration that the region is heavily doped. The basis spinor for $z<0$ takes a simple form as follows,
\begin{eqnarray}\Psi^{+(-)}_1 &=&e^{ik^{+(-)}_{1s}z}
\begin{pmatrix}
e^{-i\phi/2}&0&e^{-i\gamma^{+(-)}_F}&0
\end{pmatrix}\nonumber\\
\Psi^{+(-)}_2 &=&e^{ik^{+(-)}_{2s}z}
\begin{pmatrix}
0&e^{-i\phi/2}&0&e^{i\gamma^{+(-)}_F}
\end{pmatrix}
\end{eqnarray}
and similarly for $z>0$ the spinor are reads as,
\begin{eqnarray}\Psi^{+(-)}_3 &=&e^{ik^{+(-)}_{3s}z}
\begin{pmatrix}
e^{i\phi/2}&0&e^{i\gamma^{+(-)}_F}&0
\end{pmatrix}\nonumber\\
\Psi^{+(-)}_4 &=& e^{ik^{+(-)}_{4s}z}
\begin{pmatrix}
0&e^{i\phi/2}&0&e^{-i\gamma^{+(-)}_F}
\end{pmatrix}
\end{eqnarray}
where,
\begin{eqnarray*}
\gamma^{+(-)}_F=-\cos^{-1}\frac{\mathcal{E}^{+(-)}}{\Delta}
\end{eqnarray*}
The wavevectors $k^{+(-)}_{js}$ in superconducting region are obtained from Eq.(\ref{Eigen-BdG}). We now calculate ABSs using the method discussed in Sec-III. It is evident from Eq.(\ref{we-fflo}) and Eq.(\ref{wv-fflo}) that in the short junction limit, $k^{+(-)}_{3(4)}=k^{+(-)}_{2(1)}$ and hence $\mathcal{P}^{+(-)}_{3(4)}=\mathcal{P}^{+(-)}_{2(1)}$. These equalities are valid both in Case-I and Case-II. The ABSs spectrums are follows,
\begin{eqnarray}
E^{+(-)}_{ABS}=\Delta \sqrt{1-\Gamma^{+(-)} \sin^2\frac{\phi}{2}}
\end{eqnarray}
where the transmission probability ($\Gamma$) is given by,
\begin{eqnarray}
\Gamma^{+(-)}=\frac{(\mathcal{P}^{+(-)}_1-\mathcal{P}^{+(-)}_2)^2}{(\mathcal{P}^{+(-)}_1-\mathcal{P}^{+(-)}_2)^2+4\mathcal{P}^{+(-)}_1\mathcal{P}^{+(-)}_2 \sin^2(\Delta kL)}\nonumber\\
\end{eqnarray}
with,
\begin{eqnarray}
\Delta k=\frac{(k^{+(-)}_2-k^{+(-)}_1)}{2}=\frac{\sqrt{\mu^2_N+(C^2-1)k^2_p}}{(C^2-1)}
\end{eqnarray}
Here, $C$ is the absolute value of $C_{+(-)}$. One can check that $\Gamma^{+}=\Gamma^{-}$ for both in Case-I and Case-II. Hence, the ABSs spectrums of two chirality sectors are degenerate (i.e., $E^+_{ABS}(\phi)=E^-_{ABS}(\phi)$). This is not surprising since the FFLO pairing involves electrons at the same Weyl node. Using Eqs.(\ref{Joseph-curr},\ref{Joseph-dens}) we calculate the Josephson current. The total Josephson current from two chirality sectors is given by,
\begin{eqnarray}
J&=&(J_++J_-)\sin\phi \nonumber\\
&=&2J_0 \sin\phi 
\end{eqnarray}
with $\phi$ is the phase difference between two superconductors. $J_+$ and $J_-$ are the Josephson current contriboution from positive and negative chirality sectors, respectively. However, $J_+=J_-$ in case of FFLO pairing.


\subsection{BCS-like Pairing}
We execute similar calculations for the BCS-like pairing. We calculate ABSs and Josephson current for $H^+_B$ and $H^-_B$ here, separetly. The electron and hole wavefunctions of Hamiltonian in Eq.(\ref{bcs-final-hamil}) for the normal region are read,
\begin{eqnarray}
\Psi^{e+}_{in(out)}=e^{iq^+_{1(2)}z}\begin{pmatrix} 1 & \mathcal{Q}_{1(2)} & 0 & 0 \end{pmatrix}\nonumber\\
\Psi^{h+}_{in(out)}=e^{iq^+_{3(4)}z}\begin{pmatrix} 0 & 0 & 1 & \mathcal{Q}_{3(4)} \end{pmatrix}
\end{eqnarray}
Here, $+$ sign in uperscript corresponds to the Hamiltonian $H^+_B$. The elements $\mathcal{Q}_i$ are follows,
\begin{eqnarray}
\mathcal{Q}_1=\frac{q_pe^{i\theta}}{q^{+}_{+}+q^{+}_1};
\mathcal{Q}_2=\frac{q_pe^{i\theta}}{q^{+}_{-}+q^+_2}\nonumber\\
\mathcal{Q}_3=\frac{q_pe^{-i\theta}}{q'^{+}_{+}+q^+_3};
\mathcal{Q}_4=\frac{q_pe^{-i\theta}}{q'^{+}_{-}+q^+_4}
\end{eqnarray}
 where $q_p=\sqrt{q^2_x+q^2_y}$ is the conserved transversed momenta and $\theta=\tan^{-1}(q_y/q_x)$. The wavevectors $q^{+}_i$'s are obtained by solving eigenvalues equations of electron and hole Hamiltonian separetly, which are given,
\begin{eqnarray}
q^{+}_{1(2)}&=&\frac{C_{+}(E+\mu_N)\mp \sqrt{(E+\mu_N)^2+(C^2_+-1)q^2_p}}{(C^2_+-1)}\nonumber\\
q^{+}_{3(4)}&=&\frac{C_{-}(\mu_N -E)\pm \sqrt{(E-\mu_N)^2+(C^2_--1)q^2_p}}{(C^2_--1)}\nonumber\\
\label{wave-vector-bcs}
\end{eqnarray}
with,
\begin{eqnarray*}
q^{+}_{+(-)}&=&\sqrt{{q^+}^2_{1(2)}+q^2_p}\nonumber\\
q'^+_{+(-)}&=&\sqrt{{q^+}^2_{3(4)}+q^2_p}
\end{eqnarray*}
The quasiparticles energy spectrums are obtained by diagonalizing Eq.(\ref{bcs-final-hamil}) and given (we consider $\mu_s$ is large),
\begin{eqnarray}
\mathcal{E}^+&=&\frac{1}{2}[(C_+-C_-)q_z\nonumber \\&&\pm \sqrt{4\Delta^2+(q_z(C_++C_-\pm2)-2\mu_s)^2}]\nonumber\\
\label{quasi-spect}
\end{eqnarray}
In Case-II, Eq.(\ref{quasi-spect}) takes the following form,
\begin{eqnarray}
\mathcal{E}^+=\pm\sqrt{\Delta^2+(\mu_s-Cq_z\pm q_z)^2}
\end{eqnarray}
which are same for FFLO pairing in Eq.(\ref{Eigen-BdG}) with large $\mu_s$ limit. The tilt in this case shift the doping level $\mu_s$, with the replacement: $\mu_s \rightarrow \mu_s-C q_z$. In Case-I, Eq.(\ref{quasi-spect})  takes the following form,
\begin{eqnarray}
\mathcal{E}^+=Cq_z\pm\sqrt{\Delta^2+(\mu_s\pm q_z)^2}
\end{eqnarray}
In this case the BdG Hamiltonian has tilted energy spectrums along the $q_z$-direction.
The basis spinors for $z<0$ are follows,
\begin{eqnarray}\Psi^{+}_1 &=&e^{iq^+_{1s}z}
\begin{pmatrix}
e^{-i\phi/2} & 0 & e^{-i\gamma^+_B} & 0
\end{pmatrix}\nonumber\\
\Psi^+_2 &=&e^{iq^+_{2s}z}
\begin{pmatrix}
0&e^{-i\phi/2}&0&-e^{i\gamma^+_B}
\end{pmatrix}
\end{eqnarray}
and similarly, for $z>0$ are follows,
\begin{eqnarray}\Psi^+_3 &=& e^{iq^+_{3s}z}
\begin{pmatrix}
e^{i\phi/2}&0&e^{i\gamma^+_B}&0
\end{pmatrix}\nonumber\\
\Psi^+_4 &=&e^{iq^+_{4s}z}
\begin{pmatrix}
0&e^{i\phi/2}&0&-e^{-i\gamma^+_B}
\end{pmatrix}
\end{eqnarray}
where $\gamma^{+}_B=-\cos^{-1}(\mathcal{E}^{+}/\Delta)$ for Case-II and $\gamma^{+}_B=-\cos^{-1}((\mathcal{E}^{+}-Cq_z)/\Delta)$ for Case-I. The wave vectors $q^+_{is}$ are obtained from Eq.(\ref{quasi-spect}).
Now, in short junction limit the electron and hole wavevectors in Eq.(\ref{wave-vector-bcs}) are related: $q^{+}_{3(4)}=q^{+}_{2(1)}$ in Case-II, which are similar to FFLO pairing. In Case-I, the quasiparticles wavevectors are related: $q^{+}_{3(4)}=-q^{+}_{1(2)}$. Thus the Cooper pairs at Fermi surface have finite momentum i.e, $|q_{1(2)}-q_{3(4)}|$ is finite, while it is zero for inversion breaking tilt (Case-II) or in FFLO pairing. The finite momentum of Cooper pairs introduces an extra phase in the ABSs spectrums, which in turn leads to anomalous CPRs. However, from the above discussions it is clear that the CPRs of BCS pairing with inversion breaking tilt and FFLO pairing would be similar.

In Case-I, the analytical form of ABSs are as follows,
\begin{eqnarray}
\mathcal{E}^+_{ABS}=\Delta\sqrt{\frac{\mathcal{C}}{\mathcal{A}}-\frac{\mathcal{B}}{\mathcal{A}}\sin^2\phi^+_B}
\label{ABSp-BCS}
\end{eqnarray}
in which the expression of $\mathcal{A}$, $\mathcal{B}$ and $\mathcal{C}$  are given by,
\begin{eqnarray}
\mathcal{A}&=&(\mathcal{Q}_1\mathcal{Q}_3+\mathcal{Q}_2\mathcal{Q}_4)\cos(\Delta q L)-(\mathcal{Q}_2 \mathcal{Q}_3+\mathcal{Q}_1 \mathcal{Q}_4)\nonumber\\
\mathcal{B}&=&(\mathcal{Q}_1-\mathcal{Q}_2)(\mathcal{Q}_3-\mathcal{Q}_4)\nonumber\\
\mathcal{C}&=&(\mathcal{Q}_1\mathcal{Q}_3+\mathcal{Q}_2\mathcal{Q}_4)\cos^2(\frac{\Delta qL}{2})\nonumber\\ &&+(\mathcal{Q}_1\mathcal{Q}_2+\mathcal{Q}_3\mathcal{Q}_4)\sin^2(\frac{\Delta qL}{2})\nonumber\\&&-(\mathcal{Q}_1\mathcal{Q}_4+\mathcal{Q}_2\mathcal{Q}_3)
\label{exp}
\end{eqnarray}
with $\Delta q=(q^+_1-q^+_2)/2$. We skip $Cq_z$ term in Eq.(\ref{ABSp-BCS}) since it does not contribute the Josephson current. The expression of phase $\phi^+_B$ is given,
\begin{eqnarray}
\phi^+_B &=&\frac{(q^+_1+q^+_2)L}{2}-\frac{\phi}{2}\nonumber\\&&=\frac{\phi_t}{2}-\frac{\phi}{2}
\label{phase-BCS}
\end{eqnarray}
where the extra phase is solely due to the tilt and given $\phi_t=2\mu_N C L/(C^2-1)$. Note that, $\phi_t$ is zero for BCS pairing with inversion breaking tilt and FFLO pairing. The ABS for $H^-_B$ is obtained by replacing $\mathcal{Q}_1 \leftrightarrow \mathcal{Q}_3$ and $\mathcal{Q}_2 \leftrightarrow \mathcal{Q}_4$ in Eqs.(\ref{ABSp-BCS},\ref{exp}) and $\phi_t \rightarrow -\phi_t$ in Eq.(\ref{phase-BCS}). The total Josephson current is now given by,
\begin{eqnarray}
J&=&J_{+}+J_{-}\nonumber\\
&&=J^+_{0}\sin(\phi-\phi_t)+J^-_{0}\sin(\phi+\phi_t)
\label{total-curr}
\end{eqnarray}
We also define the current $J_{diff}$, which is given by,
\begin{eqnarray}
J_{diff}&=&J_{+}-J_{-}\nonumber\\
&&=J^+_{0}\sin(\phi-\phi_t)-J^-_{0}\sin(\phi+\phi_t)
\label{diff-curr}
\end{eqnarray}
$J_+$ and $J_-$ are the Josephson current from the BdG Hamiltonian $H^+_B$ and $H^-_B$, respectively. The current $J_{diff}$ is zero for FFLO pairing since $J_{+}$ and $J_{-}$ are equal.

We define an operator $\mathcal{D}=-i\sigma_y\mathcal{R}_y\mathcal{K}$ which relates $H^e_B$ and $H^h_B$ in Eq.(\ref{bcs-final-hamil}): $\mathcal{D}H^e_B\mathcal{D}^{-1}=H^h_B$ with $\mathcal{R}_y$ is the reflection operator about $xz$ plane and $\mathcal{K}$ is the complex conjugation.  Consequently, the symmetry of BdG Hamiltonian $\mathcal{D}_{BdG}H^{+}_{BdG}(\phi)\mathcal{D}^{-1}_{BdG}=H^{-}_{BdG}(-\phi)$ lead to the following symmetry in ABSs spectrums: $\mathcal{E}^+_{ABS}(\phi)=\mathcal{E}^-_{ABS}(-\phi)$. Here $\mathcal{D}_{BdG}=diag\{\mathcal{D},-\mathcal{D}\}$. Therefore, we have $J_+(\phi)=-J_-(-\phi)$ in this case. As a result, $J(\phi)$ is an odd (i.e., $J(\phi)=-J(-\phi)$) and $J_{diff}(\phi)$ is an even ($J_{diff}(\phi)=J_{diff}(-\phi)$) function of $\phi$, respectively. So, $J(\phi)$ vanishes always at $\phi=n\pi$ and interestingly, $J_{diff}(\phi)$ can exists even if $\phi=0$.
\begin{figure}
\includegraphics[width=1.65in]{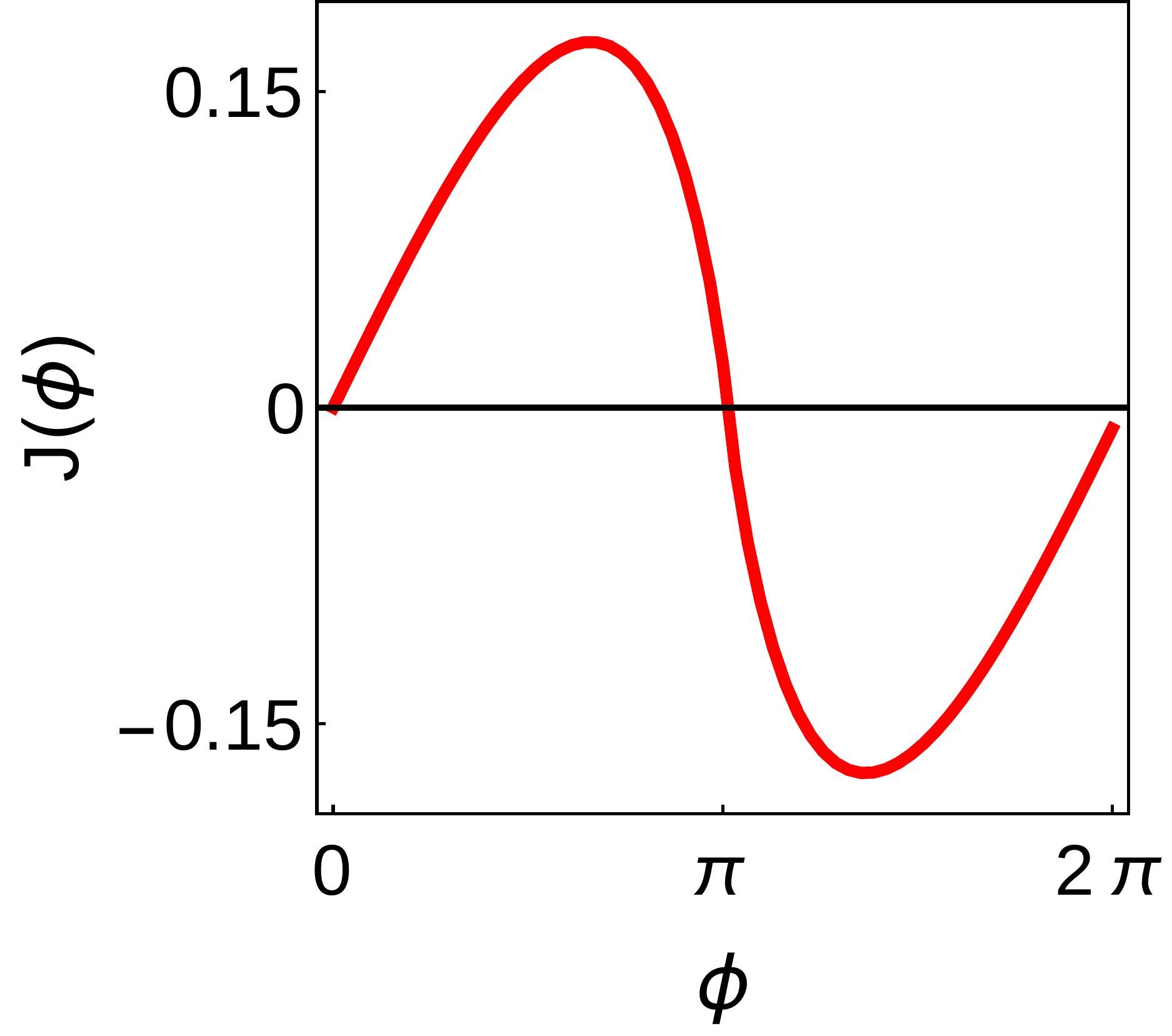}
\includegraphics[width=1.65in]{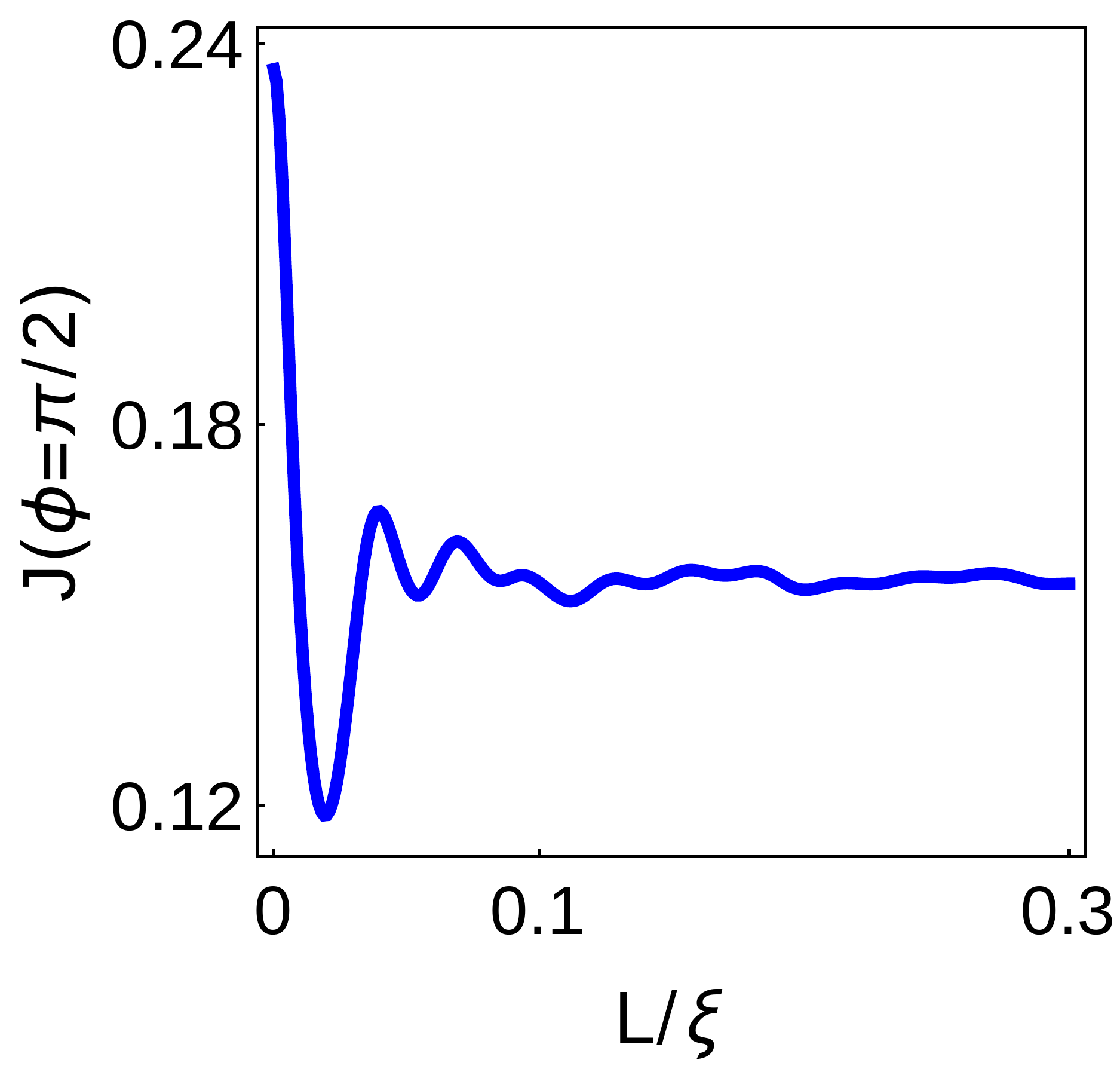}
\caption{FFLO-like pairing: Left panel displays the Josephson current as a function of superconducting phase difference $\phi$. We fix $L/\xi=0.015\pi$. Right panel displays the Josephson current with length at the phase difference $\phi=\pi/2$. In both panels, $C=0.3$ and $\mu_N/\Delta=100$. The currents are expressed in unit of $e^2\mu^2_N W^2\Delta/\hbar$.}
\label{fig1-fflo}
\end{figure}
\begin{figure}
\includegraphics[width=.3\linewidth]{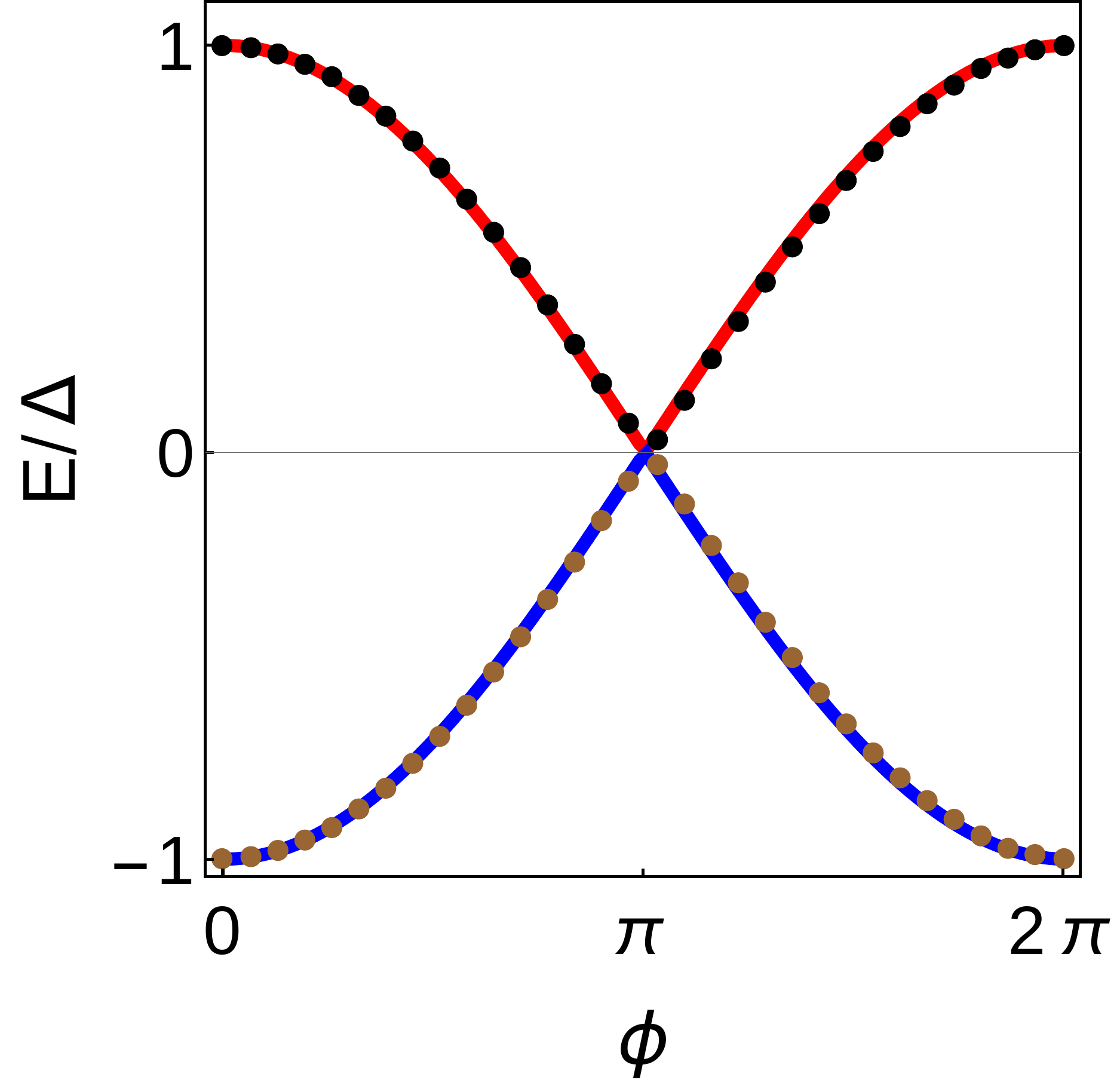}
\includegraphics[width=.3\linewidth]{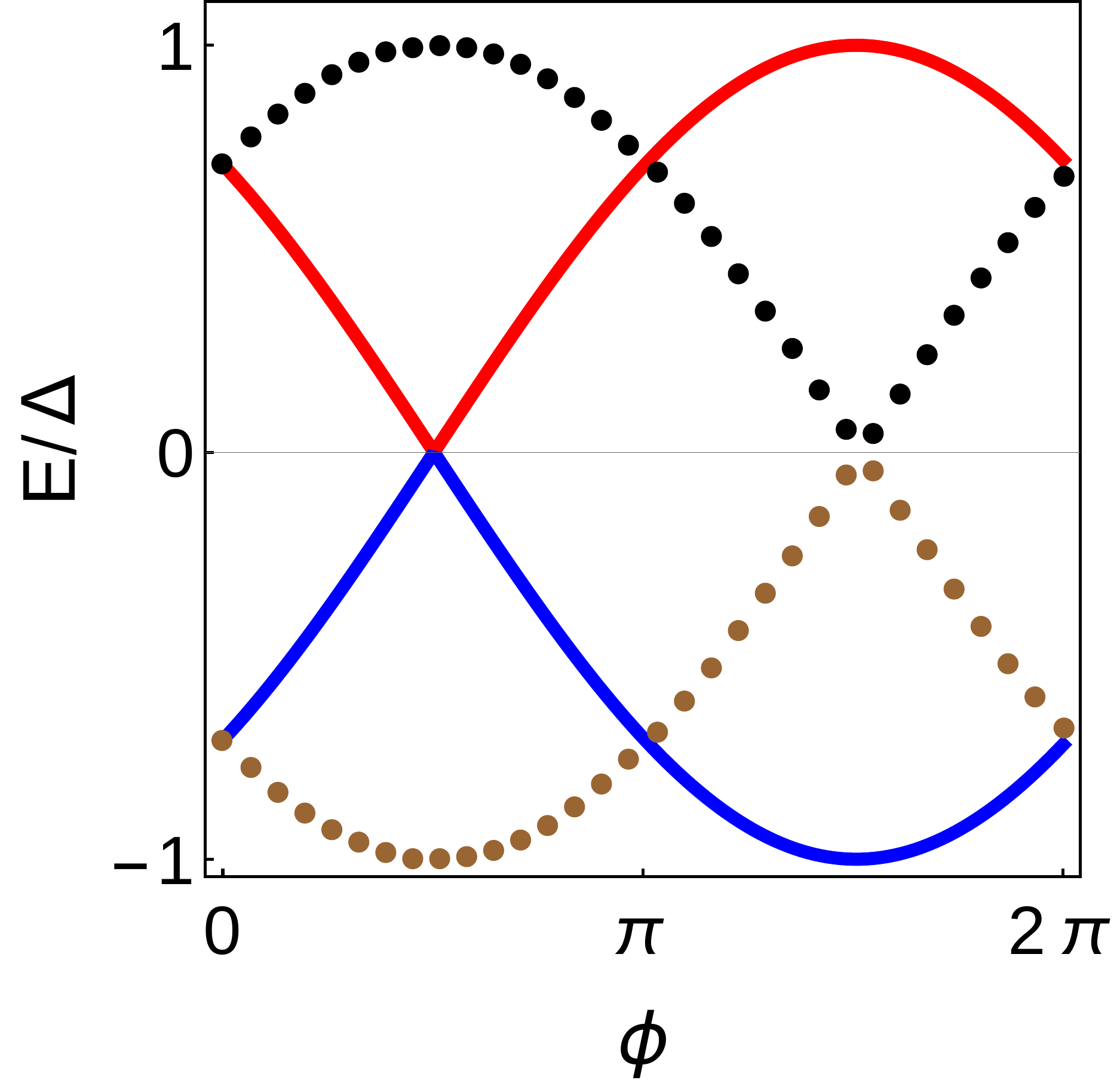}
\includegraphics[width=.3\linewidth]{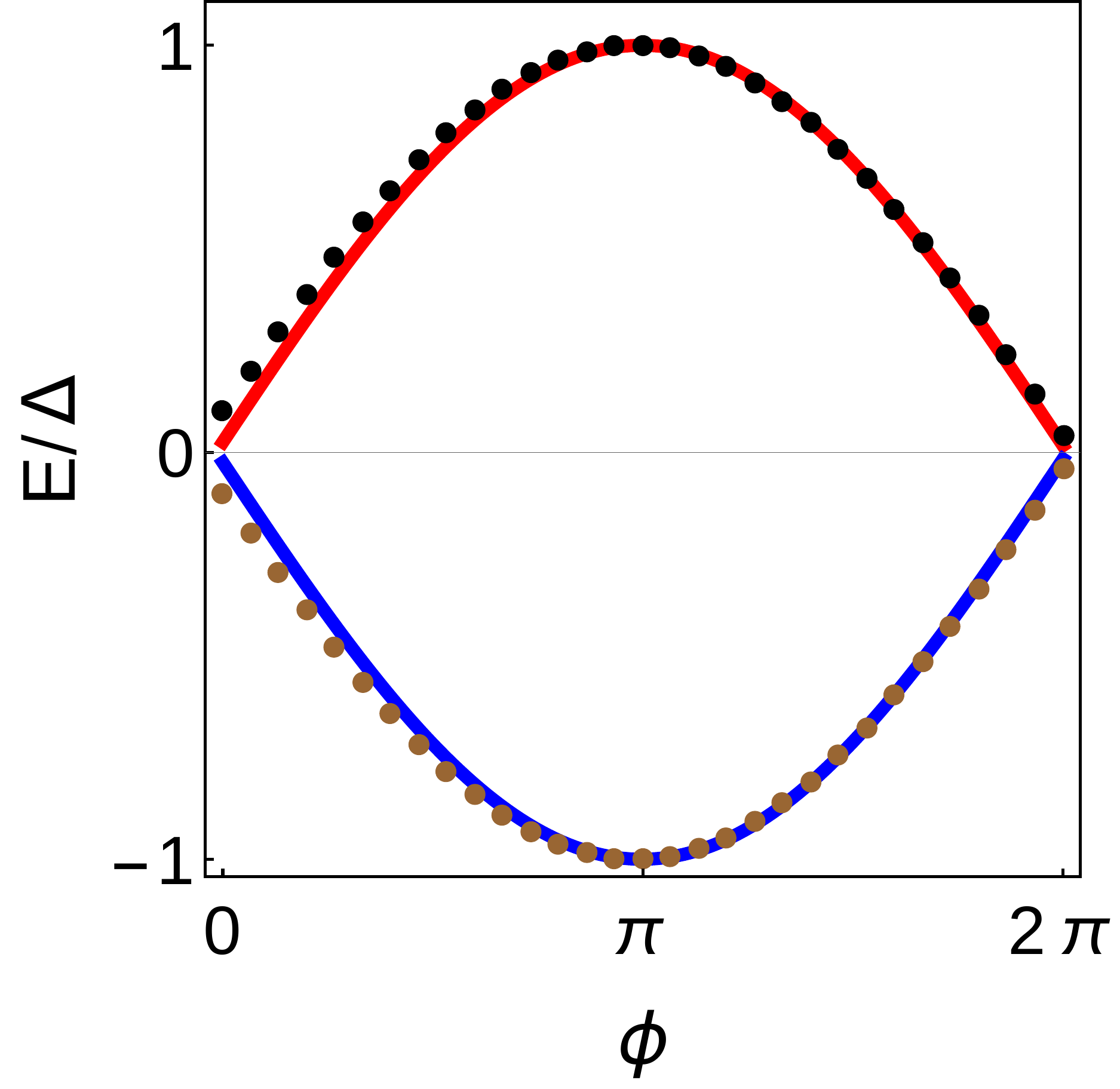}
\caption{ABSs of BCS state: Andreev Bound state as a function of superconducting phase difference $\phi$ of a type-I WSM with inversion symmetric tilt. The left, middle and right panels are corresponding to $\phi_t=2n\pi$, $(2n+1)\pi/2$ and $(2n+1)\pi$ respectively. Here, we take $C=0.3$. The red and blue lines are positive and negative ABS spectra of $H^+_B$. The black and brown dots are corresponding ABS spectra of $H^-_B$. We fix $\mu_N/\Delta=100$ and let $q_p\rightarrow 0$.}
\label{fig-ABS}
\end{figure}

\section{Results and Discussions}
The numerical results of Josephson current with $\phi$  for FFLO pairing are shown in Fig.(\ref{fig1-fflo}). In the left panel, we have shown the Josephson current variation with phase $\phi$. In the right panel, we have shown the Josephson current at $\phi=\pi/2$ with $L/\xi$. The total current follows the relation $J \sim J_0\sin\phi$ in which $J_0$ is only depends on the length and tilting parameter. Therefore, the junction always behaves as a $0$-junction.

\begin{figure}
\includegraphics[width=1.65in]{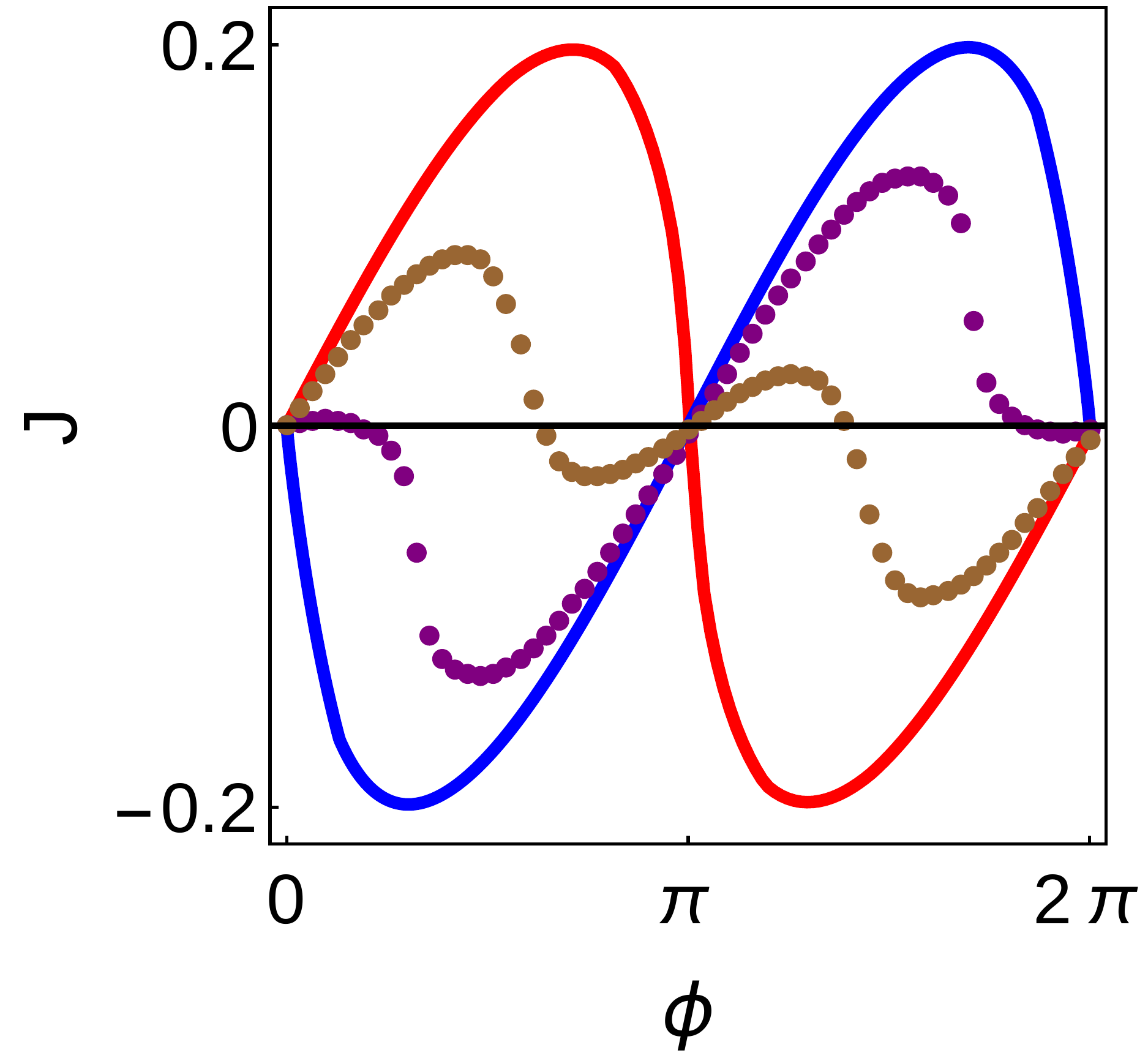}
\includegraphics[width=1.65in]{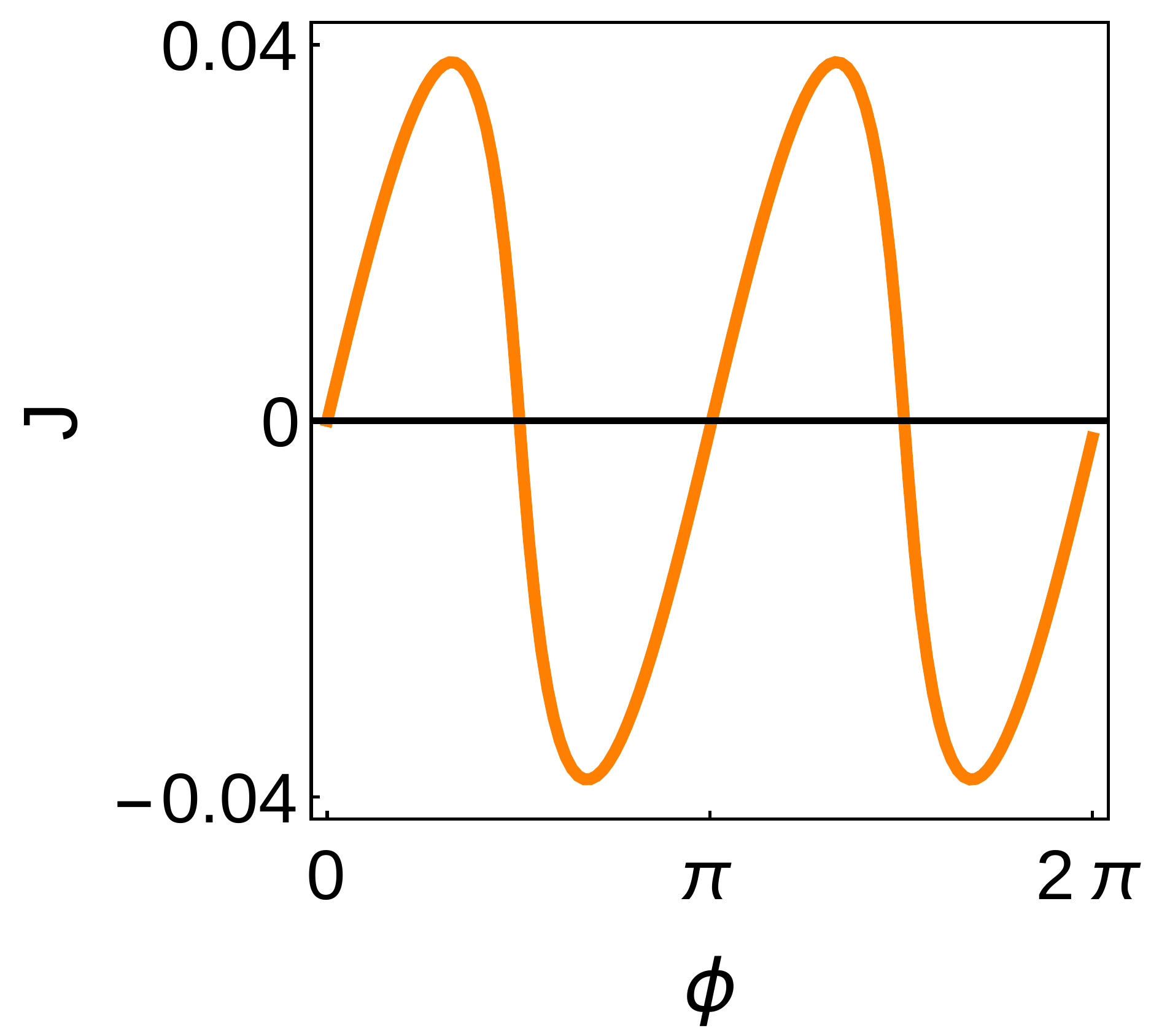}
\caption{CPR: Josephson current as a function of superconducting phase difference $\phi$. In the left panel, red and blue solid lines are corresponds to $L/\xi=0.03\pi$ and $L/\xi=0.015\pi$, which satisfies $\phi_t=2n\pi$ and $\phi_t=(2n+1)\pi$ respectively. The brown and magenta dotted lines are displays for $L/\xi=0.024\pi$ and $L/\xi=0.02\pi$, respectively. The right panel shows CPR for $L/\xi=0.0075\pi$, satisfying $\phi_t=(2n+1)\pi/2$. We fix $C=0.3$, $\mu_N/\Delta=100$. The currents are expressed in unit of $e^2\mu^2_N W^2\Delta/\hbar$. Show text for details.}
\label{fig1-bcs}
\end{figure}

To understand the non-trivial features for BCS state in Case-I, we will first discuss the ABSs. We have shown the numerical results of phase dependency of ABSs  as a parameter of $L/\xi$ with a fixed $C=0.3$ in Fig.(\ref{fig-ABS}). The ABSs for two sectors $H^+_B$ and $H^-_B$ are degenerates with the two conditions i.e., if $\phi_t=2n\pi$ or $\phi_t=(2n+1)\pi$. With the first condition, the positive root of ABS spectrums have negative (positive) slope for $\phi \in [0,\pi] (\phi \in [\pi,2\pi])$. We have shown this in the extreme left panel of Fig.(\ref{fig-ABS}). With the second condition the corresponding spectrums have positive (negative) slope for $\phi \in [0,\pi] (\phi \in [\pi,2\pi])$. We have shown this in the extreme right panel of Fig.(\ref{fig-ABS}). The corresponding Josephson current remains positive and negative for $\phi \in [0,\pi]$ with these conditions, which represents the $0$ and $\pi$-junction (shown in Fig.(\ref{fig1-bcs})), respectively. On the other hand, if $\phi_t$ (and hence $L/\xi$) is not satisfing the above conditions, then the spectrums have mixture of positive and negative slope for $\phi \in [0,\pi]$ or $\phi \in [\pi,2\pi]$. We have illustrates this in the middle panel of Fig.(\ref{fig-ABS}) for $\phi_t=(2n+1)\pi/2$. The crossing points in ABSs are shifted oppositely in the $\phi$-plane, resulting two minima at $\phi \neq 0 (\pi)$ in the spectrum within one period of $\phi$. With increasing (decreasing) the value $\phi_t$ further, the junction shifted toward the $\pi$ ($0$) and $0$ ($\pi$) junction periodically. The junction neither in $\pi$ nor in $0$ state under the conditions $\phi_t\neq 2n\pi, (2n+1)\pi$.


From the above discussions, we found that the ABS slope can be positive, negative or a mixture of these two. These spectrums correspond to the  Josephson $0$, $\pi$ and $\phi$ junction, respectively. When $\phi_t=2n\pi$, the supercureent $J(\phi)\sim \sin\phi$, which is a feature of the $0$ state. When  $\phi_t=(2n+1)\pi$, the supercureent $J(\phi)\sim -\sin\phi$, which is a feature of the $\pi$ state. So, by tuning $\phi_t$ the supercurrent $0$-$\pi$ transition can be realized. In Fig.(\ref{fig1-bcs}) we have shown the numerical results of the Josephson currents. The red and blue solid line in the left panel of Fig.(\ref{fig1-bcs}) represents Josephson $0$ and $\pi$ junction respectively. Increasing $\phi_t$ from $0$ to $\pi/2$ (or decreasing $\phi_t$ from $\pi$ to $\pi/2$) the maximum of $J(\phi)$ decreases monotonically. Also, the slope of the Josephson current at $\phi=(2n+1)\pi$ changes its sign at a critical value of $\phi_t$. These are shown by dotted curves in the left panel of Fig.(\ref{fig1-bcs}). It is also seen that an extra peak/dip occurs in those curves for $\phi \in [0,\pi]$. The current phase relation changes from $J\simeq J_c\sin\phi$ to $J\simeq J_c\sin2\phi$ at $\phi_t=\pi/2$. This is illustrated in the right panel of Fig.(\ref{fig1-bcs}).

To understand above features, we write down the Josephson current into a series of different orders of harmonics: $J(\phi)=\sum_n J_n \sin n\phi + I_n \cos n\phi$, where $J_n$ and $I_n$ decreases with $n$. Since $J(\phi)$ here is an odd function of $\phi$, this implies that $I_n=0$. The leading term is then $J(\phi)\sim\sin\phi$. From Eq.(\ref{total-curr}) it is evident that for $\phi_t=\pi/2$, $J(\phi)$ becomes $\pi$ periodic (i.e., $J(\phi)=J(\phi+\pi)$) instead of $2\pi$ periodic in $\phi$. This implies $J_{2n+1}=0$ and the first order harmonic term $\sim\sin \phi$ vanishes. Consequently, the leading term becomes the second-order harmonic term: $J(\phi)\sim \sin 2\phi$. However, around the $0$-$\pi$ transition the current phase relations is: $J(\phi)=J_{c1}\sin\phi+J_{c2}\sin 2\phi$.

\begin{figure}
\includegraphics[width=1.65in]{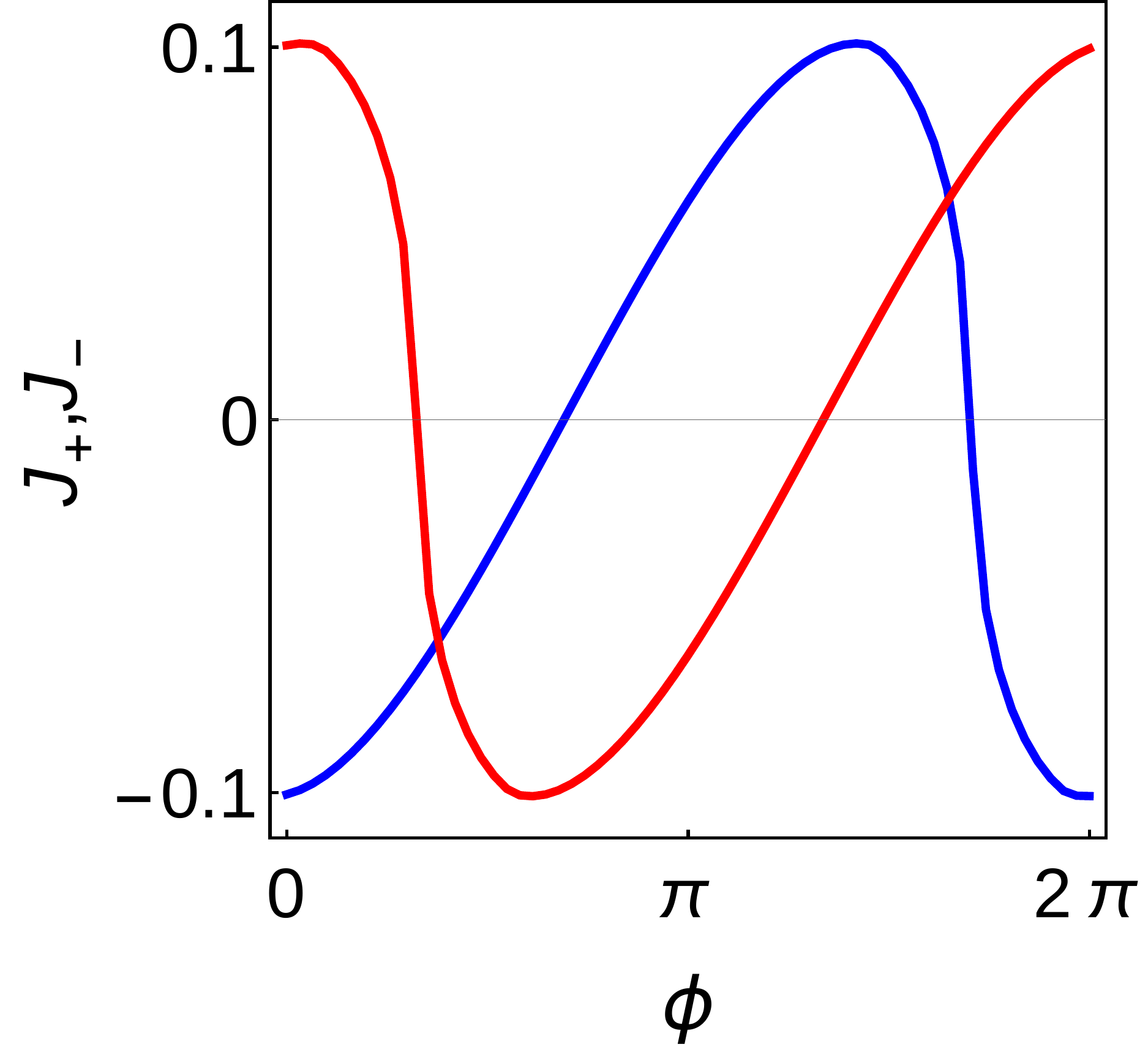}
\includegraphics[width=1.65in]{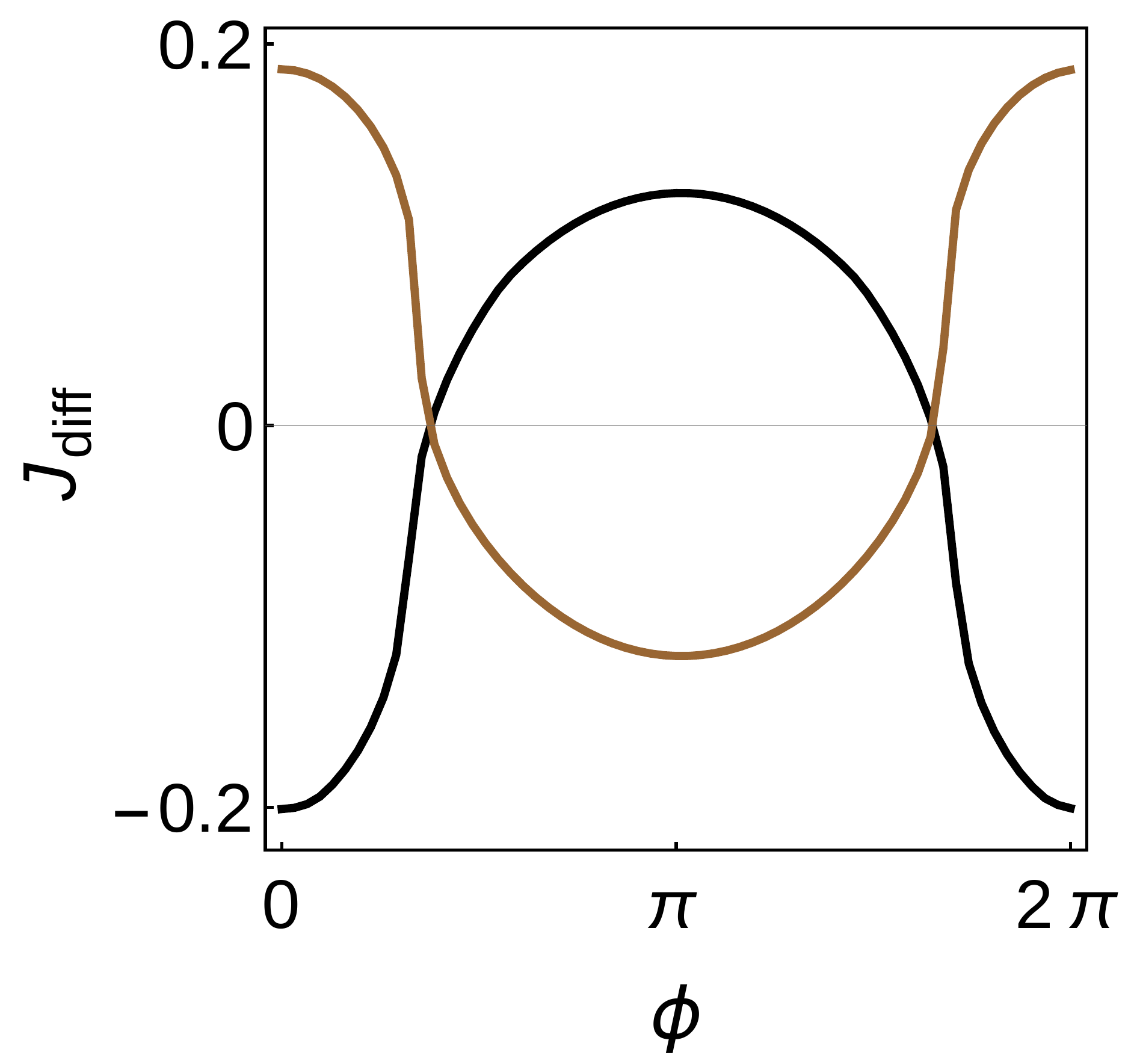}
\caption{Josephson $\phi$ junction: Left panel show the variation of $J_+$ (blue solid line) and $J_-$ (red solid line) with $\phi$ for $L/\xi=0.02\pi$, separetly. Right panel displays the difference of the Josephson currents from two sectors ($H^{\pm}_B$) as a function of superconducting phase difference $\phi$.  The black and red solid lines are for $L/\xi=0.01\pi$ and $0.02\pi$ respectively. Here we fix $C=0.3$ and $\mu_N/\Delta=100$. The currents are expressed in unit of $e^2\mu^2_N W^2\Delta/\hbar$.}
\label{fig2-bcs-diff}
\end{figure}

The Josephson current from the  two sectors $H^+_B$ and $H^-_B$ lead Josephson $\phi$ junction when either $\phi_t=2n\pi$ or $\phi_t=(2n+1)\pi$, the conditions are not satisfing. We show these in Fig.(\ref{fig2-bcs-diff}). The left panel show the variation of both $J_+$ and $J_-$ with $\phi$. The finite value of $\phi_t$ shifted the CPR between two chiralities (see Eqs.(\ref{total-curr},\ref{diff-curr})) and system indeed realize Josephson $\phi$ junction. The right panel show the variation of $J_{diff}$ with $\phi$ corresponds to the values of $\phi_t$ mentioned above. From Eq.(\ref{diff-curr}), we get $J_+=-J_0\sin\phi_t$, $J_-=J_0\sin\phi_t$ and $J_{diff}=-2J_0\sin \phi_t$ with $\phi=0$. Hence, the junction allows a finite supercurrent even if the superconducting phase difference is zero. Since $J_{diff}(\phi)$ is an even function of $\phi$, which implies it contains only $\cos n\phi$ harmonics term in the Josephson current. The leading term would be $J_{diff}(\phi)\sim \cos\phi$. So, $J_{diff}(\phi)$ has $2\pi$ periodicity in $\phi$ and the maxima of $|J_{diff}(0)|$ occurs at $\phi=n\pi$.

\begin{figure}
\includegraphics[width=1.6in]{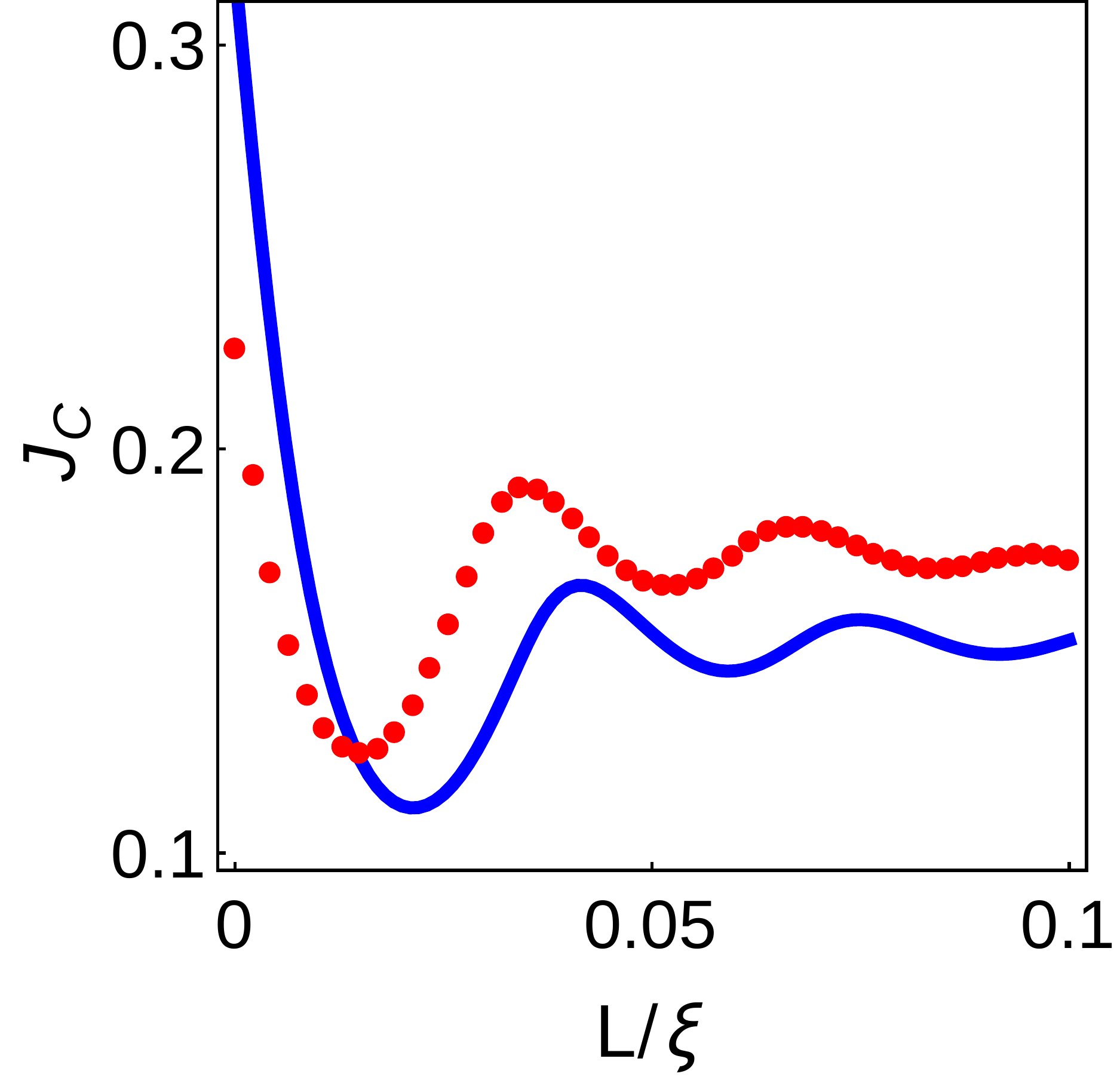}
\includegraphics[width=1.65in]{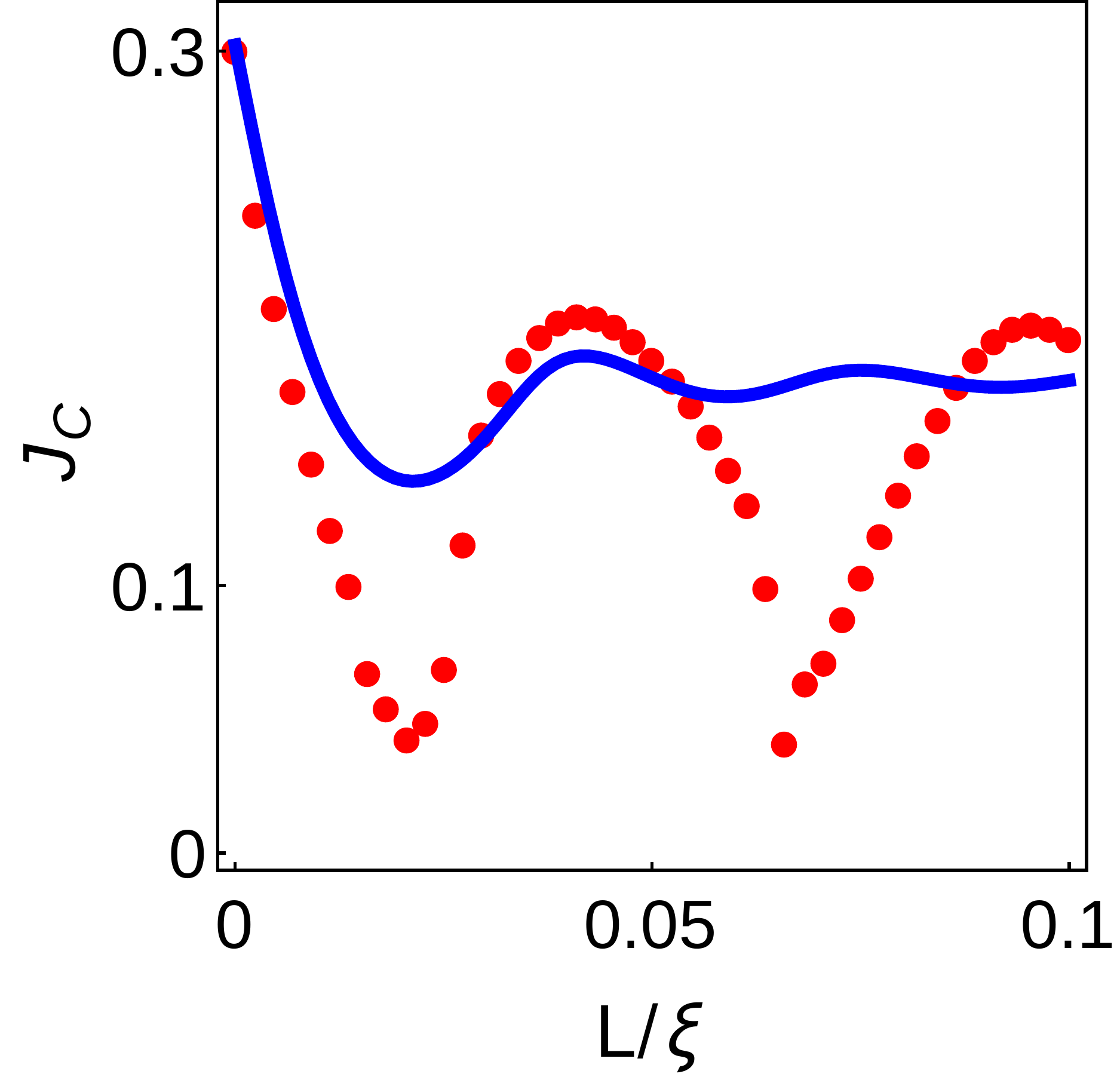}
\caption{Critical current: The critical current as a function of length $L/\xi$. The left and right panel are for FFLO and BCS -like pairing. The blue and red lines are for $C=0$ and $0.3$ respectively. Here, $\mu_N/\Delta=100$. The currents are expressed in unit of $e^2\mu^2_N W^2\Delta/\hbar$.}
\label{fig4-bcs}
\end{figure}

In order to access the experimental signature of dc Josephson current, we have shown the characteristics of junction length-dependent of critical current $J_c$ in Fig.(\ref{fig4-bcs}). The peaks in the critical current plot for BCS state give a clear indication of Josephson's current $0-\pi$ transition. For example, the right panel of Fig.(\ref{fig4-bcs}) shows a peak around $L/\xi=0.015\pi$ where the supercurrent reversal occurs (see also Fig.(\ref{fig1-bcs})). The period of oscillation of $J_c$ is compatible with the relation $J_c=2J_0\cos\phi_t$ for a finite value of $C$. The characteristics of critical current for FFLO-like pairing are shown in the left panel of Fig.(\ref{fig4-bcs}). The critical current $J_c$ oscillate and the amplitude decreases rapidly with $L/\xi$. For $C=0$, the critical current displays oscillating decays in both panels (shown by a blue solid line). The experimental studies of $J_c$ can be used to distinguish between the BCS and FFLO-like pairing in WSM.

The proximity effect of a $s$-wave superconductor on a magnetic WSM has been studied by Bovenzi et al.\cite{Bovenzi-PRB17}. They have shown that a finite interface parallel component of the vector which connects the Weyl nodes, suppressing the Josephson current from the bulk states. This phenomenon is known as 'chirality blockade'\cite{Bovenzi-PRB17}. In contrast to their work, where the superconductivity is extrinsic, we rather focused here on the intrinsic superconductivity of the doped WSMs. In our manifestation, the superconducting pairing potential of odd parity BCS state \cite{Bednik-PRB15} has pseudoscalar in nature as classified in Ref.\cite{Jafari-PRB17}. Furthermore, the superconducting gap in BCS state \cite{Moore-PRB12} is $\alpha$ dependent ($\Delta_B=|\Delta| \sin\alpha$). Here, we take $\alpha=\pi/2$ i.e., vector connecting opposite chiral nodes perpendicular to the interface. The chirality blockade will be absent in these situations, discussed in Ref.\cite{Bovenzi-PRB17,S.Jafari-PRB19}. This explains the absence of chirality blockade in our model and also in earlier studies\cite{Xing-EPL13,Madsen-PRB17}.

In the present work we restrict our study for type-I WSMs. The results can be generalized for type-II WSMs. The wavevectors of quasiparticles in Eq.(\ref{wave-vector-bcs}) are valid for type-II WSMs with $C>1$. Consequently, the BCS pairs have finite momentum and anomalous Josephson effect will continue to hold for type-II WSMs as well. However, due to the open Fermi surface and large density of states of type-II Weyl nodes, it may brings interesting physics over the type-I case. Also, in type-II WSMs there is a crtical tilt orientation at which the two nodes in the superconducting gap function disappear by merging in the Brillouin zone\cite{Alidoust-PRB17}. We will report the Josephson effects of a type-II WSM, keeping all these issues, elsewhere.

\section{Conclusions}
We investigate the Josephson effect of a TR-broken type-I WSM in a WSC-WSM-WSC junction. We consider two types of hitherto known possible pairings of a TR-broken WSM: FFLO-like and BCS-like pairing. For BCS-like pairing, a finite inversion symmetric tilt results in a wavevector shift between the electron and the hole in the Andreev Bound state. This tilt induced extra phase in wave vector modifies the phase relation of ABS and therefore leads to several remarkable features in supercurrent phase relation. We found three kinds of slopes in $\phi$-dependent ABS spectrum: positive, negative and a mixture of these two for $\phi \in [0,\pi]$. The three kinds of slopes correspond to the Josephson $0$, $\pi$ and $\phi$ junction. We demonstrate the supercurrent $0$-$\pi$ transition by tuning the parameter $\phi_t$, which means the transition can occur by tuning the length $L$ or doping $\mu_N$ for a fixed value of $C$ of a Weyl node. In the vicinity of $0$-$\pi$ transition, the first order harmonic $\sim \sin\phi$ goes to zero and second-order harmonic $\sim \sin2\phi$ becomes the leading term. However, this tilt induced phase vanishes in inversion breaking tilt. In this situation, the Josephson effect of BCS state is quite akin to the FFLO state. The junction always has a $0$-junction for FFLO-like pairing. Finally, the tilt induced exotic CPRs can provide an efficient way to understand the unconventional superconductor pairing mechanism. These anomalous CPRs realize in the absence of magnetic (Zeeman) field or unconventional superconductor in a tilted Weyl semimetal. 

\section{Acknowledgment}
We thank K. Sengupta and M. Alidoust for their useful comments and suggestions. We thank C. Beenakker for drawing our attention to the ’chirality blockade’ phenomenon reported in Ref.\cite{Bovenzi-PRB17}. We thank T. K. Bose for critically reading our manuscript and for providing useful comments.

\end{document}